# Identifying historical roots in paediatric echocardiography using RPYS

Peter Kokol[1], Jernej Završnik[2,3,4,5], Helena Blažun Vošner[2,3,6]

1. Faculty of Electrical Engineering and Computer Science, University of Maribor, Maribor, Slovenia
2. Community Healthcare Centre Dr. Adolf Drolc Maribor, Maribor, Slovenia
3. Alma Mater Europaea—ECM, Koper, Slovenia
4. Science and Research Center Koper, Koper, Slovenia
5. Faculty of Natural Sciences and Mathematics, University of Maribor, Maribor, Slovenia
6. Faculty of Health and Social Sciences Slovenj Gradec, Slovenj Gradec, Slovenia

**Abstract:** Echocardiography is a non-invasive diagnostic tool which can be performed on children of all ages including foetuses and new-borns, due to the fact that no radiation is involved. No historical bibliometric study or influential papers in paediatric echocardiography has been performed yet, so our study was aimed to close this gap. References Publication Years Spectrography (RPYS), more precisely CitedReferenceExplorer (CRE) software tool was employed to achieve this aim. We identified 35 influential papers and validation showed that both, the RPYS method and CRE tool performed satisfactorily.

## INTRODUCTION

Echocardiography is a non-invasive imaging tool that uses high-frequency sound waves (ultrasound) to create images of the heart, the blood vessels that lead in and out of it, and the flow of blood through those vessels and the heart's chambers. It is the first-line diagnostic technique in patients with congenital heart disease, and is the most common test used in children to diagnose or rule out heart disease and also to follow children who have already been diagnosed with a heart problem. This test can be performed on children of all ages and sizes including foetuses and new-borns due to the fact that no radiation is involved, and the procedure can be performed without any side effects (Grotenhuis and Mertens 2015). To the best of our knowledge no historical bibliometric study of influential papers in paediatric echocardiography has been performed yet, so our study was aimed to close this gap.

## METHODS

Robert K. Merton – the founder of the modern sociology of science – defined historical roots as important and influential publications in a specific research area (SRA) (Merton 1985). The number of citations SRA publications received seems to be an obvious measure to identify SRA's historical roots, at first glance. However, SRA publications might be (frequently) cited also by publications outside SRA, indicating that such publication could be influential in other SRAs, but not in the SRA in question. Additionally, the citing publications may not be indexed in bibliographic databases at all. To solve this problem, References Publication Years Spectrography (RPYS) method has been developed. One of the core software tools implementing this method is CitedReferenceExplorer (CRE;

www.crexplorer.net) (Thor et al. 2016, 2018), which we used in our study. The method has been already successfully used in different medical SRAs (Blažun Vošner et al. 2019; Kokol et al. 2021). CRE analyses references' publication years and aggregate the number of cited references over time. Like the spectra in the natural sciences, where pronounced peaks represent certain phenomena, pronounced peaks in the CRE spectrogram represent historical roots. We used the CRE tabular output to identify historical roots for the early period, and the CRE graphical output for later periods. Additionally, CRE indicators like the N_TOP10 indicator (that identifies publications which were among the 10% most cited publications over a longer period) were also used. To perform the analysis, we firstly formed two corpora of publications harvested from the SCOPUS bibliographic database and exported their metadata to the CRE. Those corpora included the publications presenting the application of echocardiography in paediatrics and the second corpora engineering papers related to echocardiography. CRE default parameters were used.

**RESULTS AND DISCUSSION**

CRE analysis resulted in 35 historical roots/influential papers presented in Figures 1 and Tables 1 and 2.

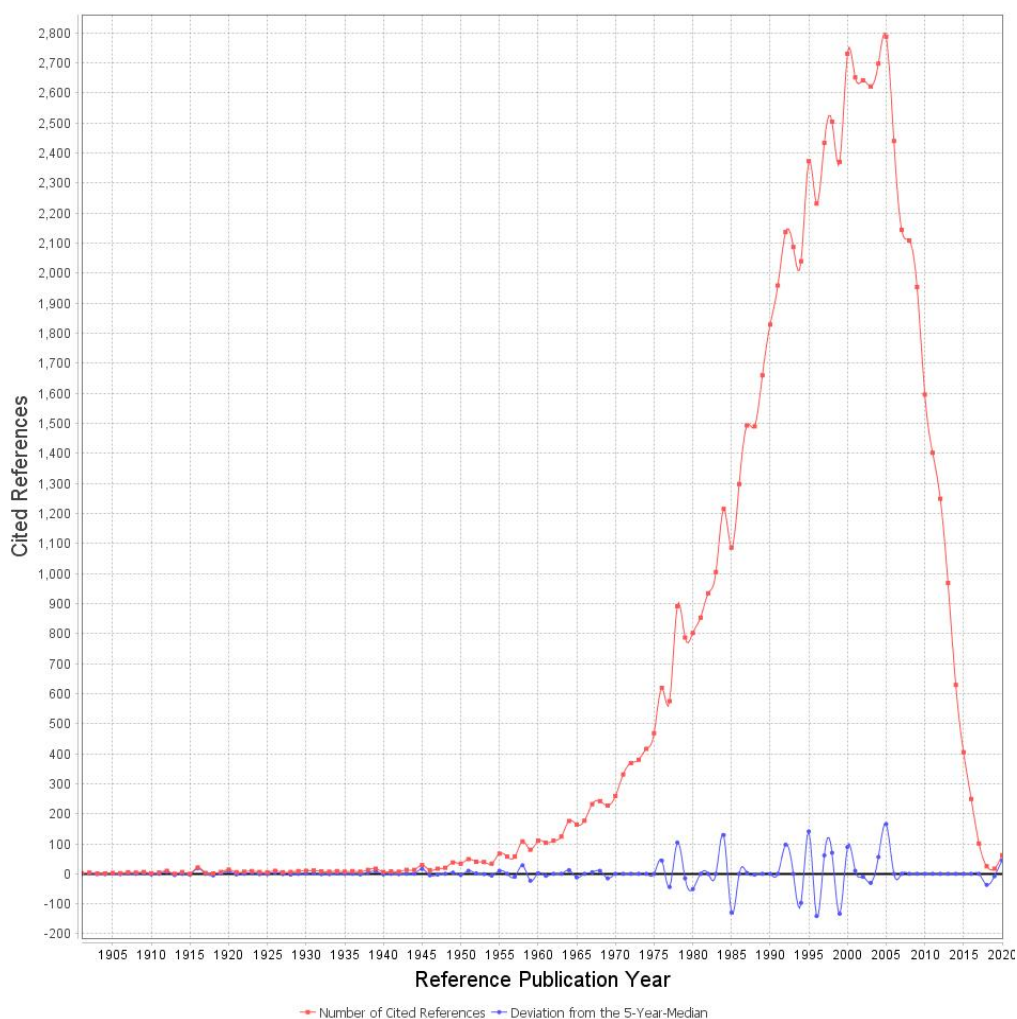

Figure 1. Paediatric historical roots in paediatrics echocardiography

Table 1. Paediatric historical roots in paediatrics echocardiography

| Pub. year | Authors | Title |
|---|---|---|
| 1628 | Harvey, W. | Exercitatio anatomica de Motu Cordis et Sanguinis in Animalibus [An anatomical disposition on the motion of the heart and blood in animals |
| 1858 | Peacock, T.B. | Malformations of the heart |
| 1866 | Ebstein, W. | Über einen sehr seltenen Fall von Insuffizienz der Valvula tricuspidalis, bedingt durch eine angeborene hochgradige Mißbildung derselben |
| 1916 | Dubois, D.; Dubois, E.F. | A formula to estimate the approximate surface area if height and weight be known |
| 1920 | Bazett, H.C. | An analysis of the time-relations of electrocardiograms |
| 1945 | Straus, R.; Merliss, R. | Primary tumor of the heart |
| 1958 | Kaplan, E.L.; Meier, P. | Nonparametric estimation from incomplete observations |
| 1976 | King, T.D.; Mills, N.L. | Secundum atrial septal defects: Nonoperative closure during cardiac catheterization |
| 1978 | Sahn, D.J.; DeMaria, A.; Kisslo, J.; Weyman, A. | Recommendations regarding quantitation in M-mode echocardiography: Results of a survey of echocardiographic measurements |
| 1984 | Colan, S.D.; Borow, K.M.; Neumann, A. | Left ventricular end-systolic wall stress-velocity of fiber shortening relation: A load-independent index of myocardial contractility |
| 1984 | Colan, S.D.; Borow, K.M.; Neumann, A. | Left ventricular end-systolic wall stress-velocity of fiber shortening relation: A load-independent index of myocardial contractility |
| 1986 | Devereux, R.B.; Alonso, D.R.; Lutas, E.M. | Echocardiographic assessment of left ventricular hypertrophy: Comparison to necropsy findings |
| 1991 | Lipshultz, S.E.; Colan, S.D.; Gelber, R.D. | Late cardiac effects of doxorubicin therapy for acute lymphoblastic leukemia in childhood |
| 1992 | De Simone, G.; Daniels, S.R.; Devereux, R.B.; Meyer, R.A.; Roman, M.J.; De Divitiis, O.; Alderman, M.H. | Left ventricular mass and body size in normotensive children and adults: Assessment of allometric relations and impact of overweight |
| 1995 | Lipshultz, S.E.; Lipsitz, S.R.; Mone, S.M. | Female sex and drug dose as risk factors for late cardiotoxic effects of doxorubicin therapy for childhood cancer |
| 1997 | Masura, J.; Gavora, P.; Formanek, A.; Hijazi, Z.M. | Transcatheter closure of secundum atrial septal defects using the new self-centering Amplatzer septal occluder: Initial human experience |
| 1998 | Eidem, B.W.; Tei, C.; O'Leary, P.W.; Cetta, F.; Seward, J.B. | Nongeometric quantitative assessment of right and left ventricular function: Myocardial performance index in normal children and patients with Ebstein anomaly |
| 2000 | Therrien, J.; Siu, S.C.; McLaughlin, P.R.; Liu, P.P.; Williams, W.G.; Webb, G.D. | Pulmonary valve replacement in adults late after repair of tetralogy of Fallot: Are we operating too late? |

Table 2. Engineering historical roots in paediatrics echocardiography

| Pub. year | Authors | Title |
|---|---|---|
| 1628 | Harvey, W. | Exercitatio anatomica de Motu Cordis et Sanguinis in Animalibus [An anatomical disposition on the motion of the heart and blood in animals |
| 1858 | Peacock, T.B. | Malformations of the heart |
| 1866 | Ebstein, W. | Über einen sehr seltenen Fall von Insuffizienz der Valvula tricuspidalis, bedingt durch eine angeborene hochgradige Mißbildung derselben |
| 1916 | Dubois, D.; Dubois, E.F. | A formula to estimate the approximate surface area if height and weight be known |
| 1920 | Bazett, H.C. | An analysis of the time-relations of electrocardiograms |
| 1945 | Straus, R.; Merliss, R. | Primary tumor of the heart |
| 1958 | Kaplan, E.L.; Meier, P. | Nonparametric estimation from incomplete observations |
| 1976 | King, T.D.; Mills, N.L. | Secundum atrial septal defects: Nonoperative closure during cardiac catheterization |
| 1978 | Sahn, D.J.; DeMaria, A.; Kisslo, J.; Weyman, A. | Recommendations regarding quantitation in M-mode echocardiography: Results of a survey of echocardiographic measurements |
| 1984 | Colan, S.D.; Borow, K.M.; Neumann, A. | Left ventricular end-systolic wall stress-velocity of fiber shortening relation: A load-independent index of myocardial contractility |
| 1984 | Colan, S.D.; Borow, K.M.; Neumann, A. | Left ventricular end-systolic wall stress-velocity of fiber shortening relation: A load-independent index of myocardial contractility |
| 1986 | Devereux, R.B.; Alonso, D.R.; Lutas, E.M. | Echocardiographic assessment of left ventricular hypertrophy: Comparison to necropsy findings |
| 1991 | Lipshultz, S.E.; Colan, S.D.; Gelber, R.D. | Late cardiac effects of doxorubicin therapy for acute lymphoblastic leukemia in childhood |
| 1992 | De Simone, G.; Daniels, S.R.; Devereux, R.B.; Meyer, R.A.; Roman, M.J.; De Divitiis, O.; Alderman, M.H. | Left ventricular mass and body size in normotensive children and adults: Assessment of allometric relations and impact of overweight |
| 1995 | Lipshultz, S.E.; Lipsitz, S.R.; Mone, S.M. | Female sex and drug dose as risk factors for late cardiotoxic effects of doxorubicin therapy for childhood cancer |
| 1997 | Masura, J.; Gavora, P.; Formanek, A.; Hijazi, Z.M. | Transcatheter closure of secundum atrial septal defects using the new self-centering Amplatzer septal occluder: Initial human experience |
| 1998 | Eidem, B.W.; Tei, C.; O'Leary, P.W.; Cetta, F.; Seward, J.B. | Nongeometric quantitative assessment of right and left ventricular function: Myocardial performance index in normal children and patients with Ebstein anomaly |
| 2000 | Therrien, J.; Siu, S.C.; McLaughlin, P.R.; Liu, P.P.; Williams, W.G.; Webb, G.D. | Pulmonary valve replacement in adults late after repair of tetralogy of Fallot: Are we operating too late? |

Echocardiography experts reviewed the roots and identified those that they didn't expect or roots that were missing. They were at first surprised by a few of the identified roots, but upon reflection they agreed that those roots are indeed important publications in the field. On the other hand, some important roots like Spallanzanis (Gacto 1999), Dopplers (Doppler and Studnica 1903), Langevin (Bok and Kounelis 2007), Dussiks (Shampo and Kyle 1995) or Edlers and Herz (Edler and Lindström 2004) publications/experiments/patents were missing. The main problem that authors didn't cite those papers in their echocardiography publications might be that they are not (yet) cited in bibliographical databases or are not published as scientific publications but as reports or patents.

**CONCLUSION**

Methodologically, RPYS supported by CRE showed to be a feasible methodology to analyse historical roots/influential papers of knowledge development in a medical SRA. In general the analysis of knowledge development can significantly contribute to the understanding of a development of a research field, to learn from past experience, and thus provide evidence for further research and enable domain experts to influence the everyday practice settings.

**FUNDING**

The study presented in this paper was partially funded by the European commission H2020 project STAMINA - Grant agreement ID: 883441